\documentclass[preprint,12pt]{elsarticle}

\usepackage{amssymb}
\usepackage{amsmath}
\usepackage{xspace}
\usepackage[euler]{textgreek}
\usepackage{hyperref}
\usepackage{booktabs}
\usepackage{subcaption}

\usepackage{lineno}

\journal{NIM A}

\newcommand{\Ricochet}{\textsc{Ricochet}\xspace}
\newcommand{\CEvNS}{CE\textnu NS\xspace}

\begin{document}

\begin{frontmatter}



\title{Results from a Prototype TES Detector for the \Ricochet Experiment}


\author[a]{C. Augier}
\author[a]{G. Baulieu}
\author[b]{V. Belov}
\author[c]{L. Berg\'e}
\author[a]{J. Billard}
\author[d]{G. Bres}
\author[d]{J-.L. Bret}
\author[c]{A. Broniatowski}
\author[d]{M. Calvo}
\author[a]{A. Cazes}
\author[a]{D. Chaize}
\author[e]{M. Chala}
\author[f]{C. L. Chang}
\author[c]{M. Chapellier}
\author[g]{L. Chaplinsky}
\author[e]{G. Chemin}
\author[h]{R. Chen}
\author[a]{J. Colas}
\author[i]{E. Cudmore}
\author[a]{M. De Jesus}
\author[c]{P. de Marcillac}
\author[c]{L. Dumoulin}
\author[d]{O. Exshaw}
\author[a]{S. Ferriol}
\author[h]{E. Figueroa-Feliciano}
\author[a]{J.-B. Filippini}
\author[j]{J. A. Formaggio}
\author[k]{S. Fuard}
\author[g]{K. Gannon}
\author[a]{J. Gascon}
\author[c]{A. Giuliani}
\author[d]{J. Goupy}
\author[e]{C. Goy}
\author[a]{C. Guerin}
\author[a]{E. Guy}
\author[j]{P. Harrington}
\author[g]{S. A. Hertel}
\author[e]{M. Heusch}
\author[i]{Z. Hong}
\author[a]{J.-C. Ianigro}
\author[l]{Y. Jin}
\author[a]{A. Juillard}
\author[b]{D. Karaivanov}
\author[b]{S. Kazarcev}
\author[e]{J. Lamblin}
\author[a]{H. Lattaud}
\author[j]{M. Li}
\author[f]{M. Lisovenko}
\author[b]{A. Lubashevskiy\footnote{Also at LPI RAS, Moscow, Russia}}
\author[c]{S. Marnieros}
\author[a]{N. Martini}
\author[j]{D. W. Mayer}
\author[d]{J. Minet}
\author[d]{A. Monfardini}
\author[a]{F. Mounier}
\author[h]{V. Novati}
\author[c]{E. Olivieri}
\author[c]{C. Oriol}
\author[h]{L. Ovalle Mateo}
\author[m]{K. J.  Palladino}
\author[g]{P. K. Patel}
\author[e]{E. Perbet}
\author[g]{H. D. Pinckney\footnote{Corresponding author: hpinckney@umass.edu}}
\author[c]{D. V. Poda}
\author[b]{D. Ponomarev\footnote{Also at LPI RAS, Moscow, Russia}}
\author[e]{F. Rarbi}
\author[e]{J.-S. Real}
\author[c]{T. Redon}
\author[j]{F. C. Reyes}
\author[e]{J.-S. Ricol}
\author[k]{A. Robert}
\author[b]{S. Rozov}
\author[b]{I. Rozova}
\author[a]{T. Salagnac}
\author[h]{B. Schmidt\footnote{Currently at CEA, Gif-sur-Yvette, France}}
\author[e]{S. Scorza}
\author[b]{Ye. Shevchik}
\author[k]{T. Soldner}
\author[j]{J. Stachurska}
\author[e]{A. Stutz}
\author[a]{L. Vagneron}
\author[j]{W. Van De Pontseele}
\author[g]{C. Veihmeyer}
\author[e]{F. Vezzu}
\author[f]{G. Wang}
\author[j]{L. Winslow}
\author[b]{E. Yakushev}
\author[f]{V. G. Yefremenko}
\author[f]{J. Zhang}
\author[b]{D. Zinatulina}

\affiliation[a]{addressline={Univ Lyon, Université Lyon 1, CNRS/IN2P3, IP2I-Lyon},
            city={Villeurbanne},
            postcode={F-69622},
            country={France}}

\affiliation[b]{addressline={Department of Nuclear Spectroscopy and Radiochemistry, Laboratory of Nuclear Problems, JINR, Dubna, Moscow Region},
            postcode={141980},
            country={Russia}}
            
\affiliation[c]{addressline={Université Paris-Saclay, CNRS/IN2P3, IJCLab, Orsay},
            postcode={91405},
            country={France}}

\affiliation[d]{addressline={Univ. Grenoble Alpes, CNRS, Grenoble INP, Institut Néel, Grenoble},
            postcode={38000},
            country={France}}

\affiliation[e]{addressline={Univ. Grenoble Alpes, CNRS, Grenoble INP, LPSC-IN2P3, Grenoble},
            postcode={38000},
            country={France}}
            
\affiliation[f]{addressline={High Energy Physics Division, Argonne National Laboratory, Lemont, IL},
            postcode={60439},
            country={USA}}

\affiliation[g]{addressline={Department of Physics, University of Massachusetts at Amherst, Amherst, MA},
            postcode={01003},
            country={USA}}
            
\affiliation[h]{addressline={Department of Physics and Astronomy, Northwestern University, Evanston, IL},
            postcode={60201},
            country={USA}}

\affiliation[i]{addressline={Department of Physics, University of Toronto, Toronto, M5S 1A7, ON},
            country={Canada}}

\affiliation[j]{addressline={Laboratory for Nuclear Science, Massachusetts Institute of Technology, Cambridge, MA},
            postcode={02139},
            country={USA}}          

\affiliation[k]{addressline={Institut Laue-Langevin, Grenoble, 38042},
            country={France}}

\affiliation[l]{addressline={C2N, CNRS, Univ. Paris-Saclay, Palaiseau},
            postcode={91120},
            country={France}}
            
\affiliation[m]{addressline={Department of Physics, University of Oxford, Oxford, UK},
            postcode={OX1 3RH},
            country={UK}}

\begin{abstract}
Coherent elastic neutrino-nucleus scattering (\CEvNS) offers valuable sensitivity to physics beyond the Standard Model. The \Ricochet experiment will use cryogenic solid-state detectors to perform a precision measurement of the \CEvNS spectrum induced by the high neutrino flux from the Institut Laue-Langevin nuclear reactor. The experiment will employ an array of detectors, each with a mass of $\sim$30~g and a targeted energy threshold of 50~eV. Nine of these detectors (the ``Q-Array'') will be based on a novel Transition-Edge Sensor (TES) readout style, in which the TES devices are thermally coupled to the absorber using a gold wire bond.  We present initial characterization of a Q-Array-style detector using a 1~gram silicon absorber, obtaining a baseline root-mean-square resolution of less than 40~eV.
\end{abstract}


\end{frontmatter}


\section{Introduction}\label{intro}

The \Ricochet experiment aims to make a detailed measurement of the coherent elastic neutrino-nucleus scattering (\CEvNS) spectrum produced by the high neutrino flux from the 58~MW Institut Laue-Langevin (ILL) nuclear reactor in Grenoble, France~\cite{augierFastNeutronBackground2023}.  An array of cryogenic detectors with a total mass of about 1~kg will be placed 8.8~m from the reactor core in a dilution refrigerator.  While the reactor at ILL can supply a high neutrino flux relative to pion decay at rest neutrino sources such as the Spallation Neutron Source at Oak Ridge~\cite{coherentcollaborationObservationCoherentElastic2017} (1.1~$\times$~10$^{12}$~cm$^{-2}$s$^{-1}$ at the \Ricochet detectors), reactor neutrino energies are lower and require a $\sim$50~eV threshold to take advantage of this high flux. To meet this requirement the \Ricochet collaboration is implementing two complementary technologies: the CryoCube and the Q-Array. 

In the baseline design of the \Ricochet experiment the CryoCube consists of 18 germanium detectors and can support an expansion of up to 27.  The detectors in the CryoCube will read out both ionization and phonon signals, and are based on optimizations of the dark matter detectors designed by the EDELWEISS collaboration~\cite{armengaudPerformanceEDELWEISSIIIExperiment2017,edelweisscollaborationOptimizingEDELWEISSDetectors2018,edelweisscollaborationFirstGermaniumBasedConstraints2020,billardSearchingDarkMatter2021}.  The phonon signal will be measured with Neutron-Transmutation Doped (NTD) sensors and ionization will be measured with electrodes and amplified by a High-Electron-Mobility-Transistor (HEMT) readout at cryogenic temperatures~\cite{billardSearchingDarkMatter2021,salagnacOptimizationPerformanceCryoCube2021}.  The combination of these two signal channels should enable the CryoCube to discriminate between electron and nuclear recoils down to $\sim$100~eV and is crucial for reducing the experimental backgrounds~\cite{colasDevelopmentDataProcessing2022}.

An alternative novel detector is proposed as the Q-Array, which utilizes a readout based on Transition-Edge Sensors (TESs)~\cite{irwinTransitionEdgeSensors2005} and radio frequency (RF) Superconducting Quantum Interference Devices (SQUIDs)~\cite{ricochetcollaborationRicochetProgressStatus2021}.  The modular design of the Q-Array, which separates the TES-based sensor chip from the absorber crystal \cite{bastidonOptimizingThermalDetectors2018,chenTransitionEdgeSensor2022,angloherFirstMeasurementsRemoTES2023}, combined with the RF-SQUID readout allows the sensors to be mass-produced, fabricated separately from the particle-sensing absorbers, and easily multiplexed.  This technology has the potential of scaling the experiment up to $\mathcal{O}(10)$~kg and beyond.  Additionally, we are studying the use of superconducting absorbers to obtain electron/nuclear recoil discrimination using pulse shape information~\cite{gaitskellNonequilibriumSuperconductivityNiobium1993}. 

In this article we present an initial characterization of a Q-Array style detector using a 1~gram silicon absorber.  This article is organized as follows: in section~\ref{fab_and_design} we describe the detector architecture and its fabrication, in section~\ref{analysis} we discuss the performance of this detector, and report the conclusions and outlook in section~\ref{conclusion}.

\section{Device Fabrication and Experimental Setup}\label{fab_and_design}

\subsection{Design Overview}\label{design}

Following the most recent Q-Array modular TES design~\cite{bastidonOptimizingThermalDetectors2018,chenTransitionEdgeSensor2022}, we implemented the first Q-Array prototype detector.  This prototype aimed to test the performance of the TES sensor chip (TES chip), described below, with a well-understood detector crystal and readout system.  Therefore we thermally coupled the TES chip to a 1~gram silicon absorber, and read out the detector with a DC SQUID system.  Details of the SQUID system are given at the end of this section.  Future effort will expand towards larger, superconducting absorbers and use RF SQUIDs to take advantage of their multiplexing capabilities.

\begin{figure}
     \centering
     \begin{subfigure}[b]{0.49\textwidth}
         \centering
         \includegraphics[width=\textwidth]{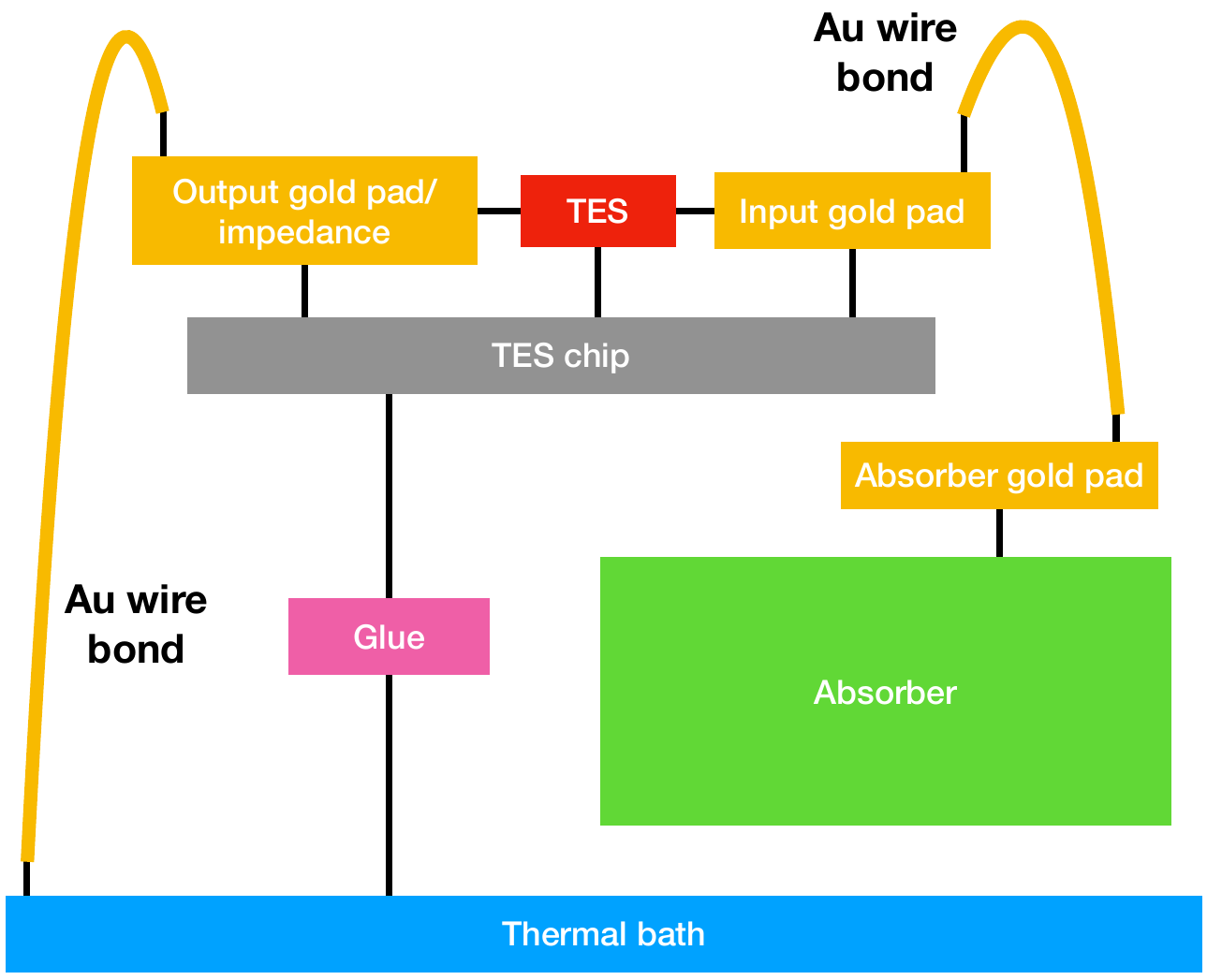}
         \caption{With absorber}
         \label{with_absorber}
     \end{subfigure}
     \hfill
     \begin{subfigure}[b]{0.46\textwidth}
         \centering
         \includegraphics[width=\textwidth]{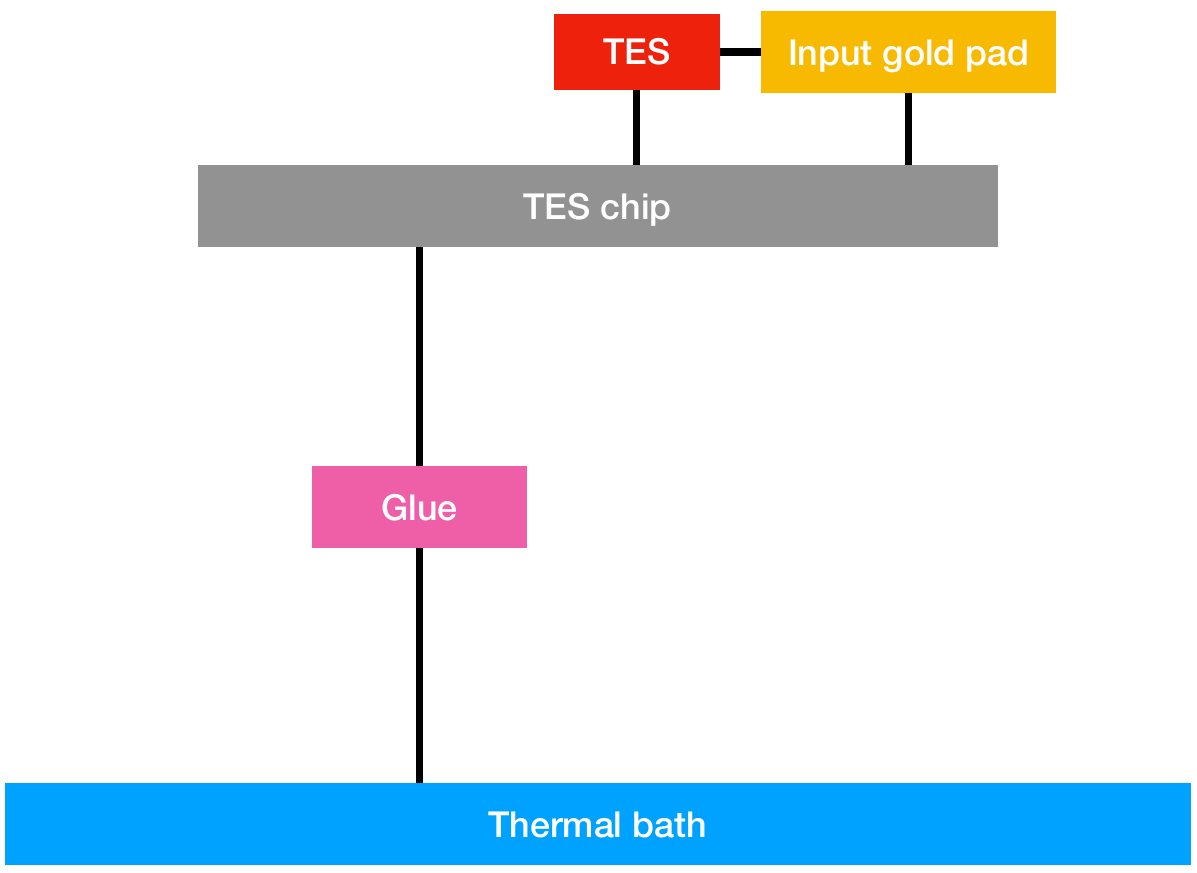}
         \caption{Without absorber}
         \label{without_absorber}
     \end{subfigure}
    \caption{Thermal block diagram of the (a) ``with absorber'' and (b) ``without absorber'' detectors under study.  The gold lines and colored blocks represent heat capacities.  These heat capacities are connected by thermal conductances, represented by black lines.  In each design the TES is read out by aluminum wire bonds to niobium electrical leads, which are not modeled in the thermal circuit.  See the text for further details.}
    \label{thermal_models}
\end{figure}

One benefit of the proposed design is the ability to engineer the heat flow into and out of the sensor.  We accomplish this through the design of the ``TES chip", a thin silicon substrate onto which we deposited the TES material, gold thermal impedance and wire bond pads, and niobium electrical leads.  In the baseline design represented in Fig.~\ref{with_absorber}, referred to as ``with absorber'' for this article, the heat flow is as follows.  The absorber and TES are connected by a gold wire bond between gold pads on both the absorber and the TES chip.  The thermal conductance of this connection is determined by the size of the absorber gold pad.  The TES and thermal bath are connected by an engineered thermal impedance and a gold wire bond, where the conductance of this connection is determined by the geometry of the impedance.  The TES chip is glued onto the thermal bath. The absorber is thermally isolated from the thermal bath via a set of sapphire balls, not illustrated, and is cooled through the TES chip.

In this article we will also discuss a ``without absorber'' detector which utilizes the silicon substrate of the TES chip as the particle absorbing target, Fig.~\ref{without_absorber}.  Heat flow into and out of the TES is controlled by the ``input gold pad", and the TES chip is thermalized to the bath through the glue.  This device serves to constrain detector parameters and acts as a check that we can distinguish events in the TES chip from events in the absorber. 

\subsection{Fabrication and Experimental Setup}\label{setup}

We fabricated the TES chips for this design at Argonne National Laboratory. All components on the TES chip, including the AlMn TES, the gold thermal impedance and wire bond pads, and the niobium leads, were patterned using non-contact optical lithography, magnetron sputtering, and the lift-off technique. The devices were fabricated on a 6~inch silicon wafer with 0.5~mm thickness and then diced into 3~mm~$\times$~3~mm chips.  This wafer contained multiple TES chip designs optimized for a variety of absorbers.  The TES is made from a two layer AlMn film with a total thickness of 200~nm.  The two layers have different levels of the Mn dopant (2000 ppm and 2500 ppm), and varying the relative film thicknesses of the two layers allows for adjusting the T$_\text{c}$ of the resulting film.  The TES surface was passivated with a thin bilayer film of 15~nm Ti and 15~nm Au to avoid oxidation. All layers were deposited at room temperature in the same vacuum chamber in a DC magnetron sputtering system with a base pressure of about 2~$\times$~10$^{-8}$~Torr.  Wafer holder rotation and tilted 3~inch sputtering guns ensured the thickness uniformity of the film.  After deposition, the chips were heated for 10~minutes at 180~$^{\circ}$C to set the T$_\text{c}$~\cite{2016JLTP..184...66L}.  We note that the ``without absorber'' TES chip was heated an additional time to 175~$^{\circ}$C for one hour prior to testing with the goal of further increasing its T$_\text{c}$.

\begin{figure}[htb]
\begin{center}
\includegraphics[width=\columnwidth]{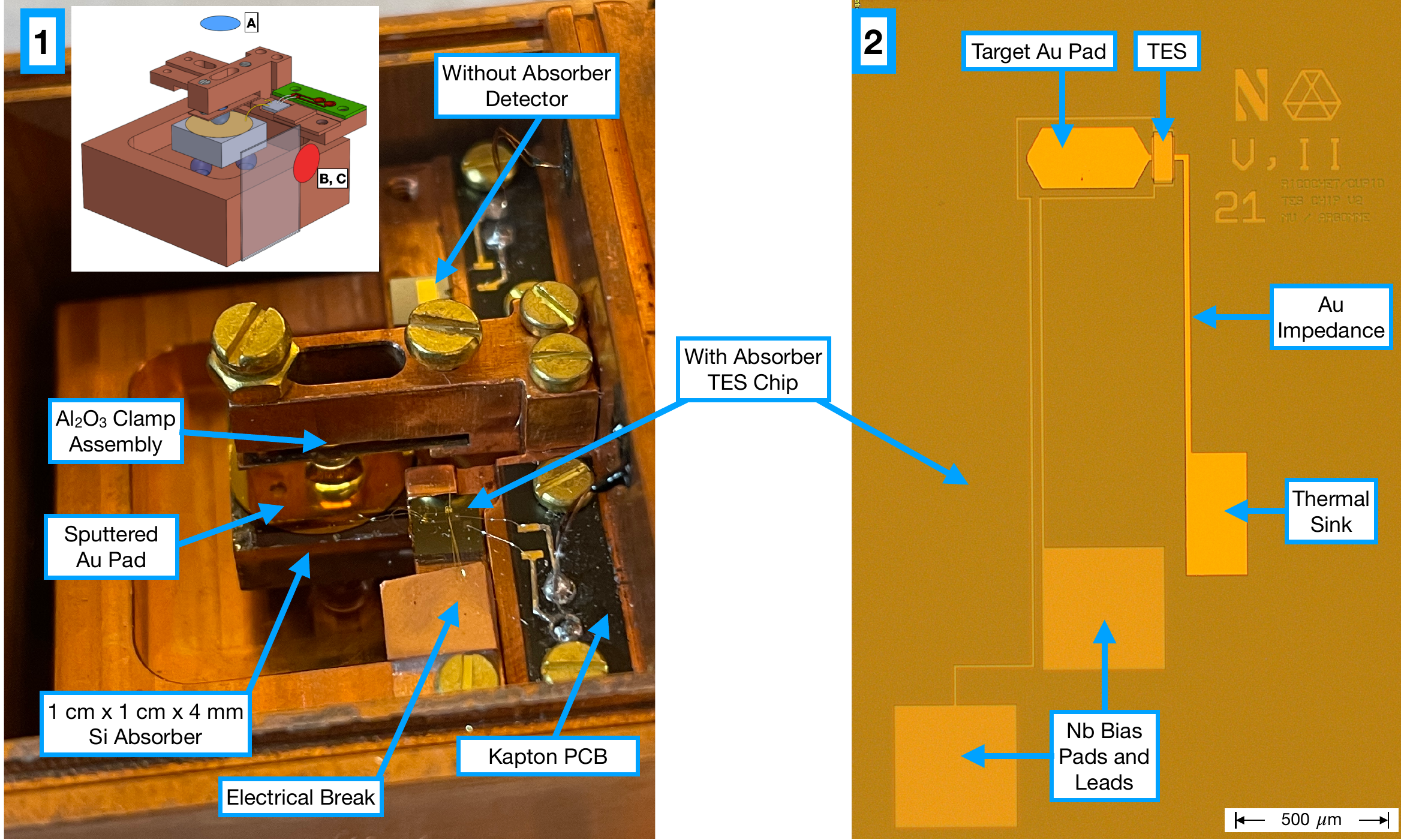}
\end{center}
\caption{Panel 1: The detector housing without its lid, showing the ``with absorber'' and ``without absorber'' TES chips.  The inset shows the source locations for our three data runs: A, B, and C.  Further details in the text.
Panel 2: The ``with absorber'' TES chip with the AlMn TES, niobium electrical leads, and gold engineered thermal impedance.  Further details in the text.}
\label{detector_setup}
\end{figure} 

We used the R\&D detector holder shown in Fig.~\ref{detector_setup} for this test.  This holder allows for simultaneous operation of two TES chips, and in this study we operate one detector ``with absorber" and another ``without absorber".  In each case the TES chip was secured to the detector holder with GE-7031 varnish, acting as the ``glue" in Fig.~\ref{thermal_models}.  The electrical contact with the TESs was made with aluminum wedge bonds to the niobium pads on the TES chip.

We coupled the TES chip on the ``with absorber'' detector to a 1~gram, 1~cm~$\times$~1~cm~$\times$~4~mm silicon absorber crystal.  A 200~nm thick, 10~mm diameter gold pad was sputtered onto this crystal.  Thermal connection between the TES chip and the silicon absorber was accomplished by two 17~$\mu$m diameter, $\sim$1~cm long gold wedge bonds.  The thermal connection between the TES chip and detector housing is made with three 17~$\mu$m diameter, $\sim$5~mm long gold wedge bonds between the chip's designed thermal impedance and a thermally-conductive electrical break which acts as a thermal bath.  The electrical break is composed of a copper-epoxy-copper sandwich, where the epoxy layer is approximately 25~$\mu$m thick Stycast 1266.  This break is necessary to electrically isolate the detector and prevent ground loops.  The absorber was thermally isolated from the detector housing with four 3.18~mm diameter sapphire balls (Swiss Jewel B3.18S), where three support the bottom of the crystal and one provides constraint from the top~\cite{pinckneyThermalConductanceSapphire2022}. 

We calibrate the detectors with X-rays from an approximately 40~Bq $^{55}$Fe X-ray source installed inside the detector box.  This source primarily provides 5.9~keV photons from K$_\alpha$ de-excitations in $^{55}$Mn.  The data we present were taken over the course of three runs with this X-ray source in locations named A, B, and C, see the inset of Panel 1 of Fig.~\ref{detector_setup}.  In run A we placed the source on the inside of the detector housing lid, facing downward onto the top of the silicon absorber.  In runs B and C we placed the source on the detector housing wall, facing the side of the silicon absorber.  In run C we added an X-ray shield made of a copper tape to obscure the top sapphire ball and most of the silicon absorber, testing for position dependent effects in the detector.

\begin{figure}
\begin{center}
\includegraphics[width=0.5\columnwidth]{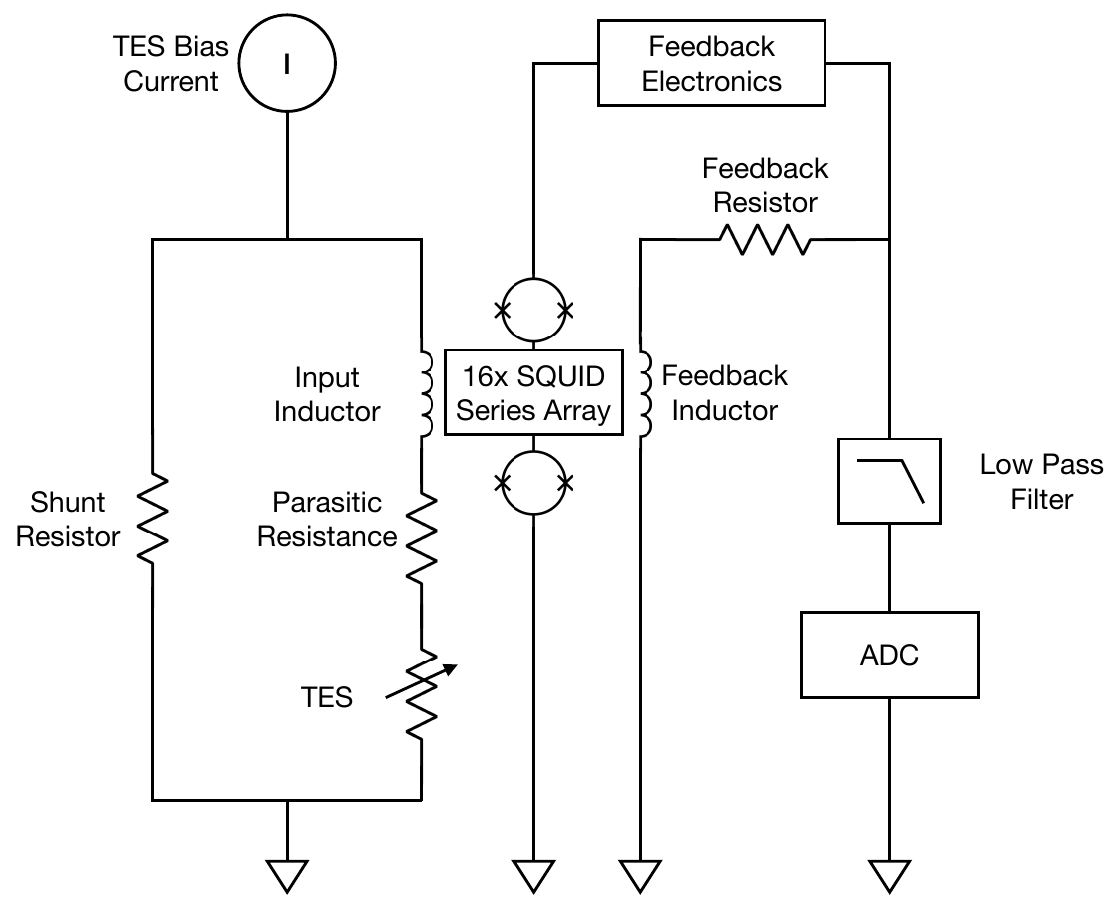}
\end{center}
\caption{Electrical circuit of our readout system.  Details presented in the text.}
\label{circuit_diagram}
\end{figure}

We cooled the setup to below 10~mK on the mixing chamber (MC) of a dilution refrigerator (Cryoconcept Hexadry UQT-B 400) at the University of Massachusetts Amherst, and our readout scheme is presented in Fig.~\ref{circuit_diagram}.  Each detector used an approximately 2~m$\Omega$ shunt resistor, and was read out with a Magnicon C6X16F SQUID series array.  The SQUIDs have a typical current noise of $\sim$6~pA/$\sqrt{\text{Hz}}$ and a ratio of ``feedback sensitivity'' over ``input coupling'' is $\sim$1.3, where feedback sensitivity and input coupling are in units of $\mu$A/$\phi_0$.  The two SQUID channels were controlled with a single Magnicon FLL XXF-1, and recorded with a National Instruments PCIe-6376 16-bit data acquisition (DAQ) card.  A 10~kHz low-pass, anti-aliasing filter was installed between the SQUID output and the DAQ input, and data was recorded continuously with a sample rate of 300~kHz.  Data was intentionally over-sampled to overcome the noise floor of the DAQ, increasing the sampling frequency from 30~kHz to 300~kHz reduces the DAQ noise from $\sim$4~pA/$\sqrt{\text{Hz}}$ to $\sim$1~pA/$\sqrt{\text{Hz}}$.

\section{Detector Characterization and Performance}\label{analysis}

\subsection{TES Properties}

\begin{figure}[htb!]
\begin{center}
\includegraphics[width=0.75\columnwidth]{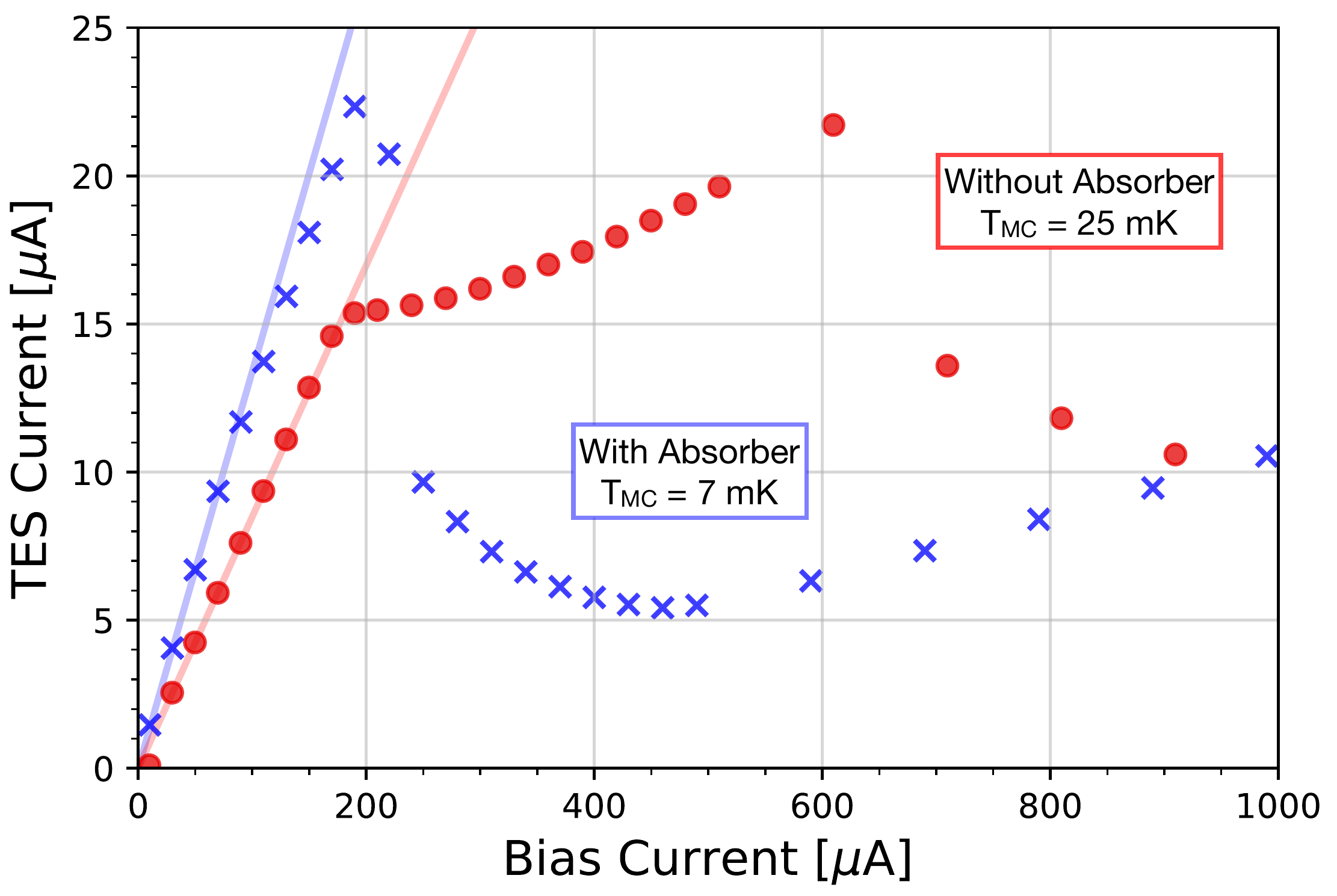}
\end{center}
\caption{Bias current vs SQUID-measured TES current (IBIS) curves.  Data are shown for the ``with absorber'' (blue crosses) and ``without absorber'' (red circles) detectors.  Due to the difference in T$_\text{c}$ the ``with absorber'' detector IBIS curve was taken with a mixing chamber temperature (T$_\text{MC}$) of 7~mK, and the ``without absorber'' detector IBIS curve was taken with a MC temperature of 25~mK.  The superconducting branch in each detector is highlighted with a faint solid line.  Each detector shows two transition features, though this feature is more dramatic in the ``without absorber'' detector.  This potentially indicates the presence of a proximity effect in the TES chips, see text for details.}
\label{iv_curve}
\end{figure}

We characterized the TES chips through their bias current (IB) vs SQUID-measured TES current (IS) characteristics, referred to here as IBIS curves.  An example is illustrated in Fig.~\ref{iv_curve}.  Lines of constant slope intersecting the origin are proportional to the inverse of the TES resistance plus parasitic resistance from wiring.  At high bias current the TES is kept normal through self-heating, and the IBIS curve slope is relatively low indicating the normal resistance.  As the bias current decreases the Joule heating power in the TES decreases and it cools down.  This decreases the resistance, increasing the slope between the data point and the origin.  At a low enough bias current the TES enters its superconducting state and the slope of the IBIS curve becomes constant again, reflecting the parasitic resistance remaining in the TES readout circuit.  In the measured IBIS curves we observe multiple transition features in each TES.  We hypothesize this is due to a combination of two proximity effects: (1) the overlap between the gold and TES (5~$\mu$m), and (2) the overlap between the niobium pads and the TES (10~$\mu$m).  The TES properties are summarized in Table~\ref{tes_properties}.

\begin{table}[htb]
    \begin{center}
        \begin{tabular}{llll}
        \toprule
        Detector & \hspace{0.5cm} R$_\text{n}$~[m$\Omega$]  & \hspace{0.5cm} T$_\text{c}$~[mK]     & \hspace{0.5cm} Bias Power~[pW] \\
        \midrule
        With Absorber   & \hspace{0.5cm} 180~$\pm$~20    & \hspace{0.5cm} 18~$\pm$~2  & \hspace{0.5cm} 4.1~$\pm$~0.4  \\
        Without Absorber & \hspace{0.5cm} 200~$\pm$~20    & \hspace{0.5cm} 38~$\pm$~4  & \hspace{0.5cm} 15~$\pm$~2      \\
        \bottomrule
        \end{tabular}
    \end{center}
    \caption{Parameters of the TES chips as determined by IBIS curve measurements.  Here, R$_\text{n}$ is the TES normal resistance, T$_\text{c}$ is the critical temperature of the TES, and bias power is the power dissipated in the TES through Joule heating when biased at 50\% of the normal resistance.  Because of the difference in T$_\text{c}$, bias power for ``with absorber'' was measured with T$_\text{MC}$~=~7~mK and bias power for ``without absorber'' was measured with T$_\text{MC}$~=~25~mK.  Uncertainties on R$_\text{n}$ and bias power are dominated by the 10\% systematic uncertainty on the shunt resistor, and uncertainty on T$_\text{c}$ is dominated by the 10\% systematic uncertainty on our MC thermometer calibration.}
    \label{tes_properties}
\end{table}

In addition, we measured the thermal conductance between each TES and the MC.  We assume that a single conductance dominates, and that this conductance can be written as a power law of the form~\cite{MatterMethodsLow}

\begin{equation}
    \text{G}(\text{T}) = \text{A}\text{T}^\beta\quad,
\end{equation}

\noindent where G is the thermal conductance, T is the temperature of the conductance, and both A and $\beta$ are constants.  Since the definition of thermal conductance is

\begin{equation}
    \text{G}(\text{T}) = \frac{\text{dP}}{\text{dT}}\quad,
\end{equation}

\noindent where P is the power through a cross-sectional plane within the thermal link, we can integrate to find that

\begin{equation}
    \text{P} = \text{K}\big(\text{T}_\text{c}^{\text{n}} - \text{T}_\text{b}^{\text{n}}\big)\quad,
\label{power_equation}
\end{equation}

\noindent where T$_{\text{b}}$ is the temperature of the thermal bath, $\text{n}=\beta+1$, and $\text{K} = \text{A/n}$. 

To take advantage of the relationship in Equation~\ref{power_equation}, we measure IBIS curves at various bath (MC) temperatures and find the Joule power required to keep the TES in transition when  biased at 50\% of the normal resistance.  We then extract the conductance between the TES and the bath by performing two fits to Equation~\ref{power_equation}: first with the exponent as a free parameter and then again with the exponent fixed. When the exponent is a free parameter it converges to 1.73 and 3.54 for the ``with absorber'' and ``without absorber'' detectors respectively.  We take this as motivation to fix the exponent in the second fits to 2 and 4 for the ``with absorber'' and ``without absorber'' detectors respectively.  For the ``with absorber'' detector this represents the assumption that the conductance is dominated by electron transport, potentially in the designed thermal impedance, and for the ``without absorber'' detector this represents the assumption that the conductance is dominated by phonon transport, potentially through the varnish securing the chip to the housing.  The choice to fix the exponent only varies the conductance within the established systematic errors.

Results from the fits are displayed in Table~\ref{conductance_table}.  We assume the measurement uncertainty is dominated by the 10\% systematic uncertainty on the shunt resistor and the 10\% systematic uncertainty on the MC temperature.  We characterize these effects by re-fitting the data at the boundaries of these uncertainties and combining their effects in quadrature, assuming they are uncorrelated.  The conductances we measure are in order-of-magnitude agreement with the expectation from our thermal model, and future work will study the details of this comparison.

\begin{table}[ht]
    \begin{center}
        \begin{tabular}{lllll}
        \toprule
        Detector           &  n        & T$_\text{c}$ [mK]          &  G [pW/K] at T$_\text{c}$         &  G [pW/K] at 18~mK   \\
        \midrule
        With Absorber      &  1.73     &  18~$\pm$~2                &  490$^{+120}_{-100}$              &  490$^{+120}_{-100}$ \\
        \addlinespace[0.5em]
                           &  2 - fixed &  18~$\pm$~2                &  540$^{+150}_{-120}$              &  540$^{+150}_{-120}$ \\
        \addlinespace[0.5em]
        Without Absorber   &  3.54     &  38~$\pm$~4                &  1700$^{+900}_{-600}$             &  300$^{+130}_{-90}$  \\
        \addlinespace[0.5em]
                           &  4 - fixed &  38~$\pm$~4                &  1900$^{+1100}_{-700}$            &  200$^{+110}_{-80}$  \\
        \bottomrule
        \end{tabular}
    \end{center}
    \caption{The thermal conductance fit parameters from the ``with absorber'' and ``without absorber" detectors, as determined by IBIS curve measurements at different MC temperatures. The values of n are first allowed to vary, and then fixed in the fit.  Results are reported for each of these conditions, as indicated in the table.  Uncertainties on T$_\text{c}$ represent the 10\% systematic uncertainty on our MC thermometer.  Uncertainties on G represent the 10\% systematic uncertainty on our shunt resistor and the 10\% systematic uncertainty on our MC thermometer, combined in quadrature under the assumption that they are uncorrelated.}
    \label{conductance_table}
\end{table}

\subsection{Detector Performance}

We study detector performance through analysis of three continuously recorded datasets, one each for runs A, B, and C.  The analysis is based on the implementation of an optimal filter (OF) as both our trigger and our energy estimator~\cite{Pytesdaq2022}.

\begin{figure}
\begin{center}
\includegraphics[width=0.75\columnwidth]{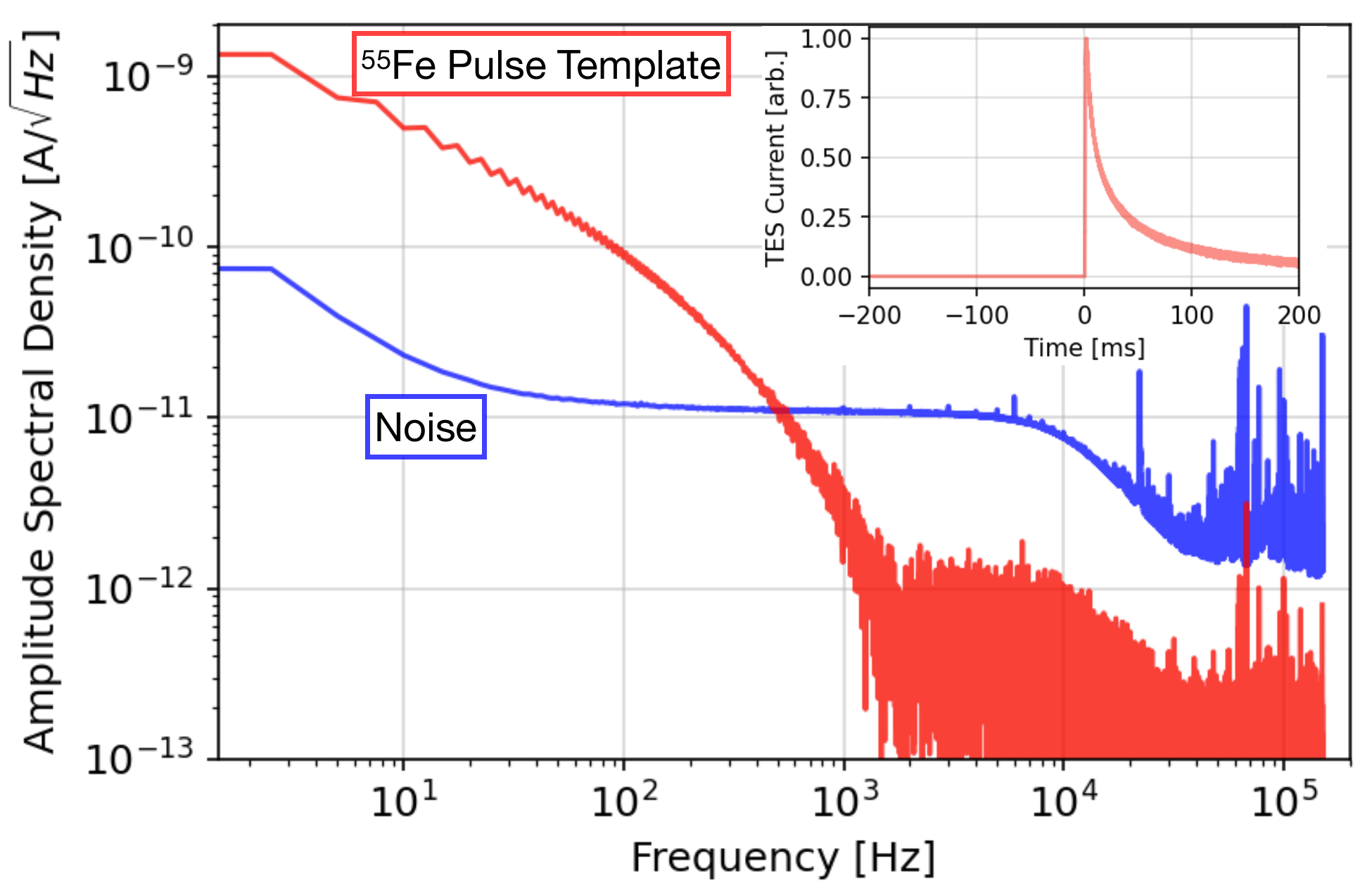}
\end{center}
\caption{Construction of the optimal filter for events hitting the silicon absorber.  Here we show amplitude spectral densities, the square root of the PSD, of the detector noise in-transition (blue) and of the pulse template scaled to 5.9~keV (red).  The inset shows the template plotted in the time domain with arbitrary normalization.}
\label{psd_template}
\end{figure}

We calculate the OF using the noise power spectral density (PSD) and the pulse template.  The noise PSD is estimated from events triggered randomly, after rejecting events with a significant slope, event baseline, poor chi-squared when compared with a pulse fit, and trace pile-up.  The pulse template is determined by averaging tens of events in the center of the 5.9~keV K$_\alpha$ peak from the $^{55}$Fe source.  We then construct an OF with this PSD and pulse template, and trigger the data with a threshold of 5 times our baseline root-mean-square (RMS) resolution.  An example of the pulse template and noise used to construct the OF is shown in Fig.~\ref{psd_template}.

We perform the energy calibration by assuming a linear scaling between 0~eV and the estimated location of the $^{55}$Fe K$_\alpha$ peak.  The peak location was estimated by performing a weighted mean on the spectrum histogram in the region containing the peak.  For each dataset the weighted mean was within 2\% of the peak center, so we assign a 2\% systematic uncertainty to the calibration.  We validated the linearity assumption by observing that the same scaling could be applied up to the Compton edge of 662~keV gamma rays from $^{137}$Cs.  

\begin{figure}
\begin{center}
\includegraphics[width=0.75\columnwidth]{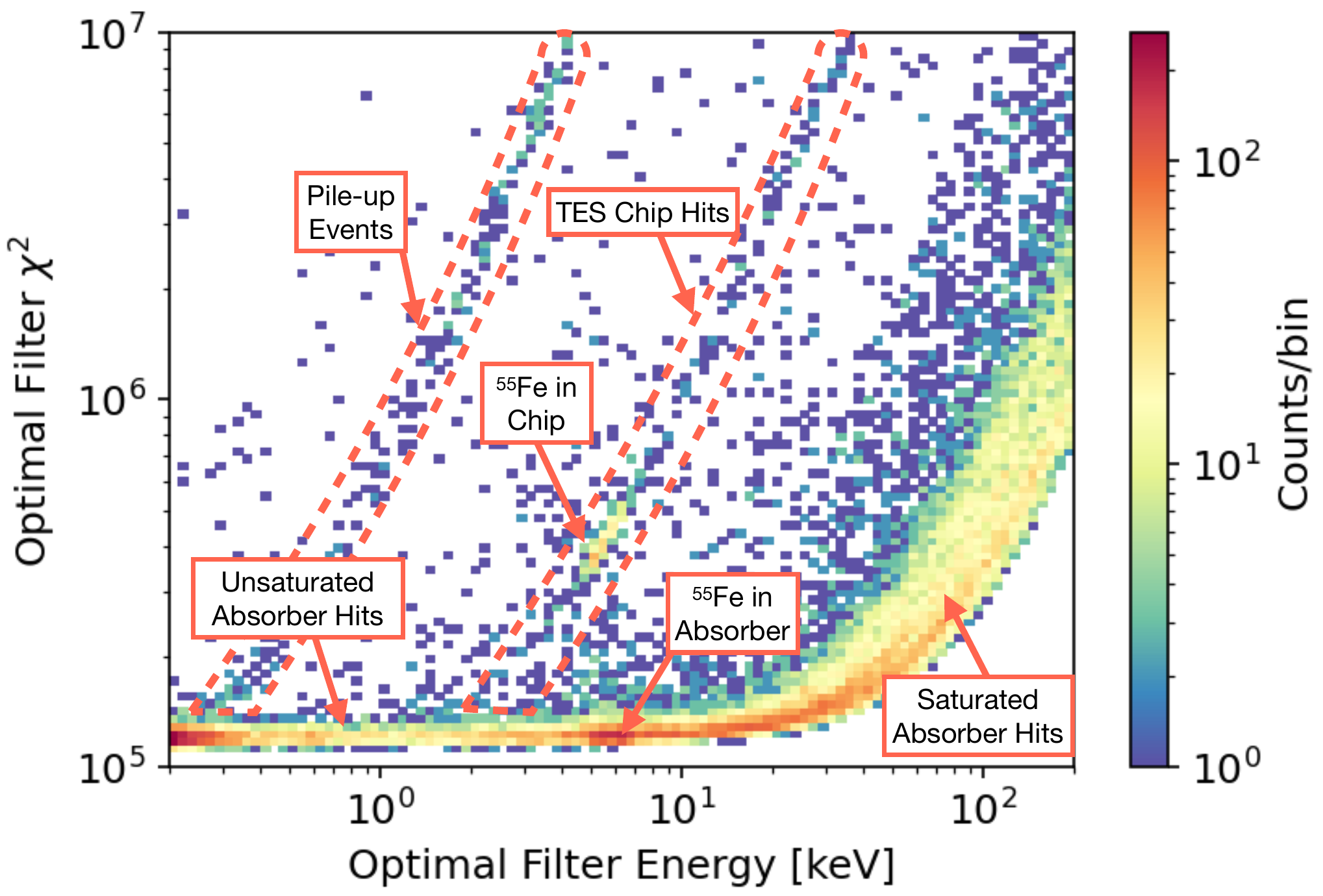}
\end{center}
\caption{The optimal filter chi-squared value vs energy parameter space for the ``with absorber'' detector during run A.  Three branches are present, identified as pile-up events, chip hits, and absorber hits.  Red dashed lines highlight the pile-up and chip hit event regions.}
\label{of_chi2}
\end{figure}

\begin{figure}
\begin{center}
\includegraphics[width=0.75\columnwidth]{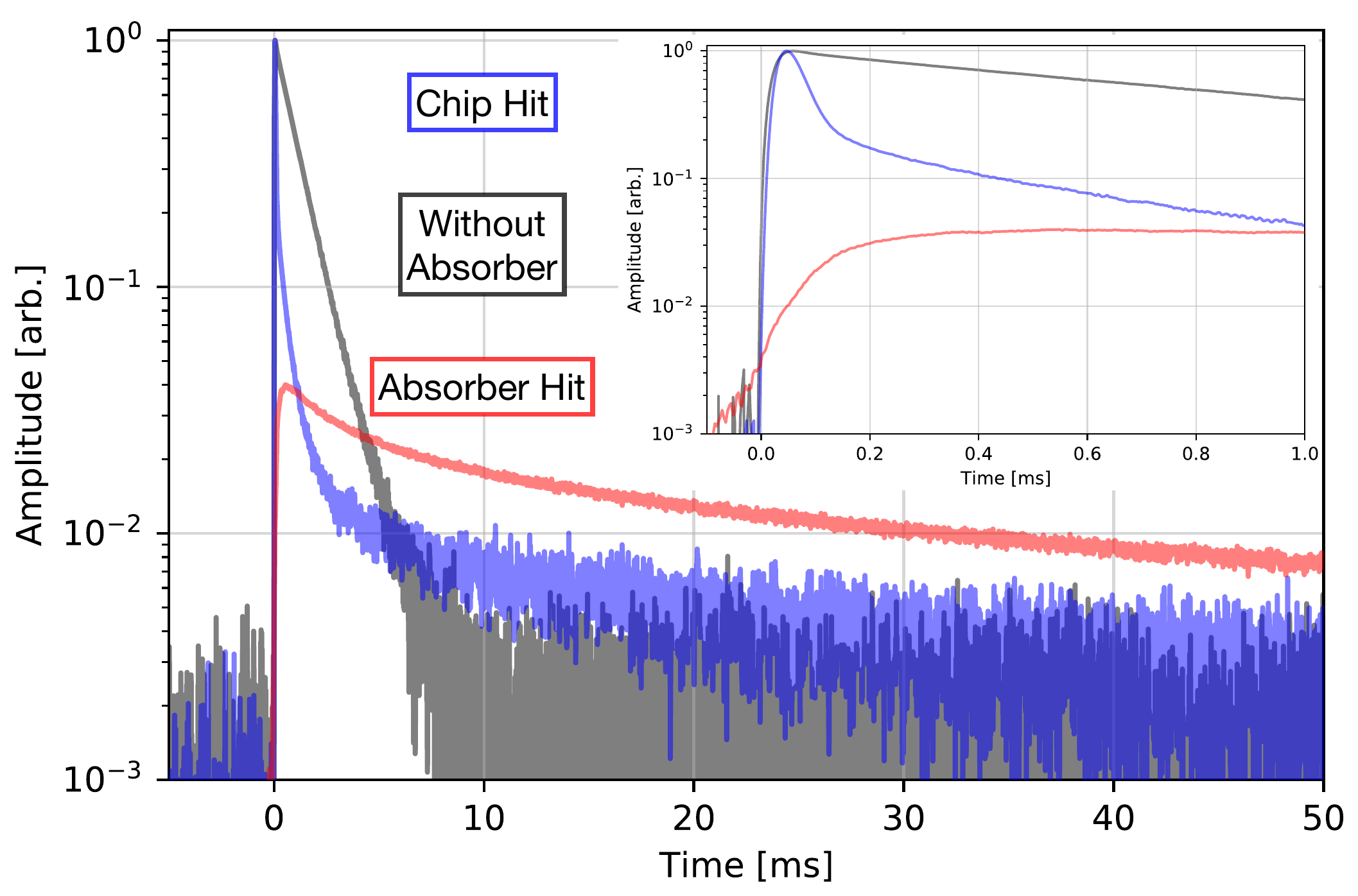}
\end{center}
\caption{Pulses from absorber hits (red), TES chip hits (blue), and the ``without absorber'' detector (black).  The inset zooms in on the first 1~ms following a trigger.}
\label{avg_pulses}
\end{figure}

The chi-squared between the data and the pulse fitted with the OF is used as an estimator of how much the event deviates from the assumed pulse template shape.  Figure~\ref{of_chi2} shows the distribution of the OF chi-squared value as a function of OF-estimated energy.  Three distinct bands are present, which we label as the ``pile-up events'', ``TES chip hits'', and ``absorber hits'' branches.  We identified the distinction between TES chip hist and silicon absorber hits through use of the ``without absorber'' detector, whose events appear similar to the TES chip hit branch on the ``with absorber'' detector. Figure~\ref{avg_pulses} shows example pulses from the TES chip hit and absorber hit branches.  Each of these pulse populations shows distinct time constants, which we currently describe with one rise time and two fall times.  The TES chip hit pulse is faster on the falling edge of the pulse, indicative of the lower heat capacity of the TES chip compared to the silicon absorber.  Chip hit events can be described by: $\tau_{\text{rise}} = $~123$\pm$8~$\mu$s, $\tau_{\text{fall-1}} = $~31$\pm$0.02~$\mu$s, $\tau_{\text{fall-2}} = $~483$\pm$2~$\mu$s, $\tau_{\text{fall-3}} = $~15.7$\pm$0.1~ms, and absorber hit events can be described by: $\tau_{\text{rise}} = $~152$\pm$0.6~$\mu$s, $\tau_{\text{fall-1}} = $~2.88$\pm$0.01~ms, $\tau_{\text{fall-2}} = $~21.54$\pm$0.08~ms, $\tau_{\text{fall-3}} = $~167.0$\pm$0.5~ms.  Detailed comparison between the measured pulse shape and the thermal model is ongoing, and will be further confirmed with more data taken.

\begin{figure}
     \centering
     \begin{subfigure}[b]{0.49\textwidth}
         \centering
         \includegraphics[width=\textwidth]{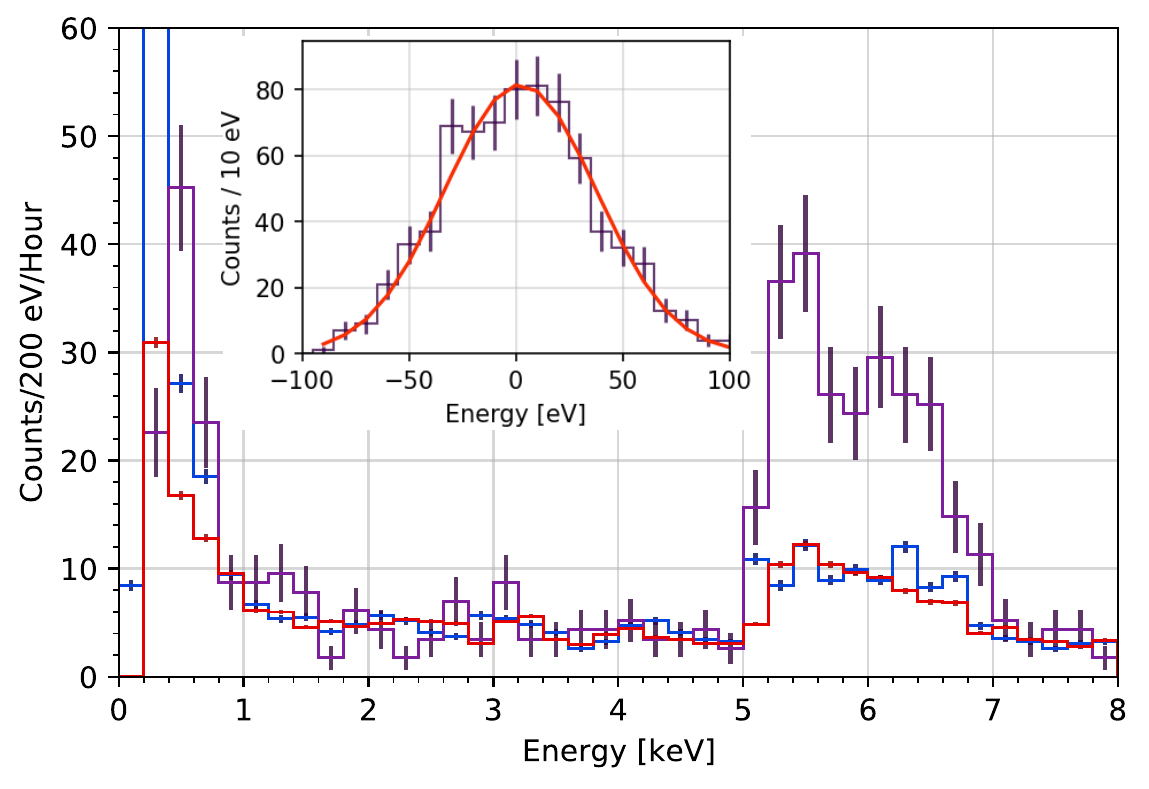}
         \caption{High energy spectra.}
         \label{high_energy}
     \end{subfigure}
     \hfill
     \begin{subfigure}[b]{0.49\textwidth}
         \centering
         \includegraphics[width=\textwidth]{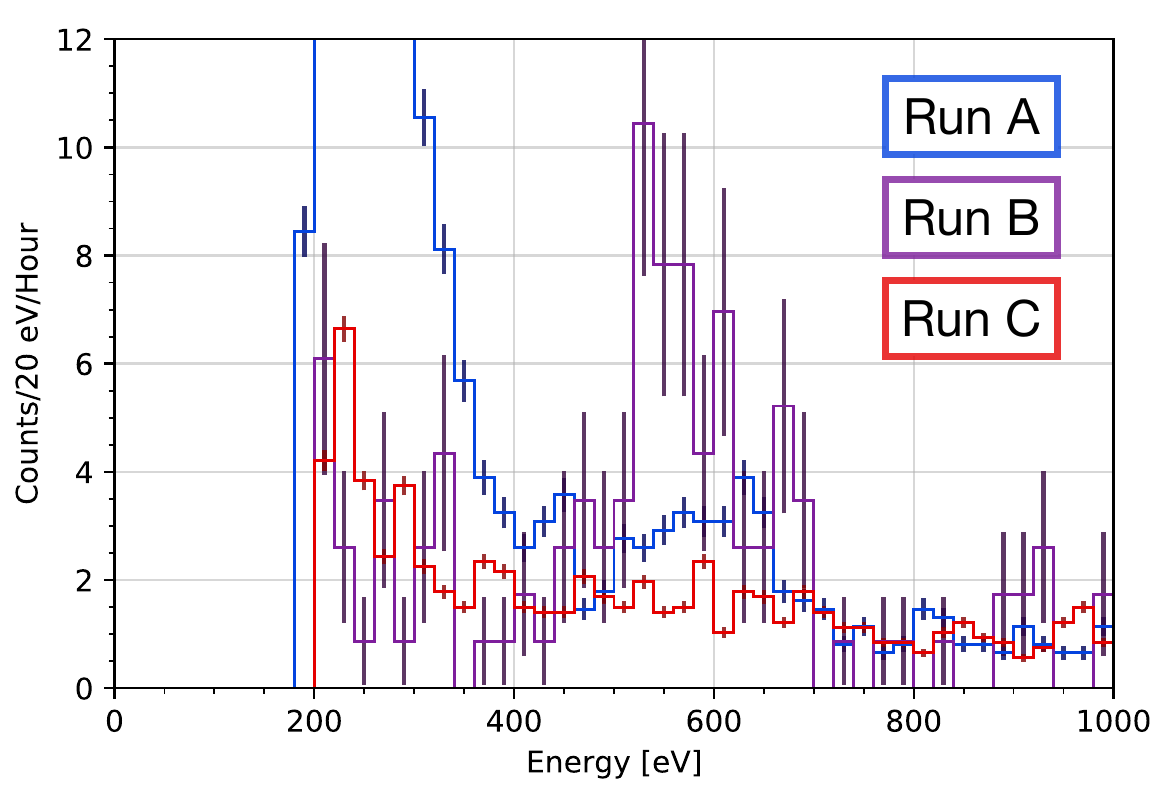}
         \caption{Low energy spectra.}
         \label{low_energy}
     \end{subfigure}
    \caption{Energy spectra in two energy regions for runs A (blue), B (purple), and C (red).  In (a) we show the spectrum of silicon absorber hits in the $^{55}$Fe event region, as well as an estimate of the baseline energy resolution in the inset.  In (b) we show the spectrum of silicon absorber events below 1 keV.}
    \label{spectrum}
\end{figure}

Figure~\ref{spectrum} shows the energy spectrum after selecting for events in the silicon absorber and applying data quality cuts.  These cuts are applied relative to iteratively estimated parameter medians and standard deviations.  For each iteration step, we re-calculate the median and standard deviation while excluding events further than one standard deviation from the median.  We begin by rejecting pileup events which occur before the pulse window by removing events whose baseline average is less than the median minus one standard deviation.  Of the events remaining we further reject pre-trace pileup by removing those whose baseline slope is greater than the median plus one standard deviation after iterating once.  Next, in order to remove times of abnormally high noise, we remove events whose baseline standard deviation is greater than the median plus one standard deviation after iterating twice.  Finally, to remove the pileup and chip-hit branches we remove events whose fit chi-squared was greater than the median plus one sigma after iterating twice.  

Following these cuts we investigate the spectrum in two regions to highlight both the $^{55}$Fe event peak, and the lower energy features in the data.  Of note in the lower energy region is a peak at approximately 550~eV in Run B.  This peak disappears when the sapphire ball is protected from line-of-sight hits from the $^{55}$Fe source with a copper X-ray shield in Run C.  This is consistent with the hypothesis that this peak is due to X-rays absorbed in the sapphire ball.  

Finally, we estimate the baseline energy resolution with a Gaussian fit to the OF-estimated amplitude distribution of randomly selected traces with pulse rejection cuts.  These cuts are performed by iteratively rejecting events greater than a specified number of standard deviations from the dataset mean, with the mean and standard deviation updated at each iteration.  The  iteration stops once the mean and standard deviation shifts less than a specified tolerance on three sequential iterations.  Specifically, we use the method implemented in~\cite{QETpy2023} to reject outliers on the baseline mean,  baseline slope, chi-squared of a pulse fit, and pile-up event amplitude, cutting on 2, 2, 3, and 2 standard deviations respectively.  To remove pileup events occurring before the event window we fit each trace for the amplitude of a decaying exponential with a fall time of 100 ms and remove any traces where this amplitude is greater than 2$\times10^{-9}$~A.  

The RMS resolution determined this way was consistently measured to be $(36~\pm~1~\text{stat.}~\pm~0.8~\text{syst.})$~eV, where the statistical uncertainty comes from the statistics of the fit, and the systematic uncertainty comes from the calibration.  This is an order of magnitude better than the observed resolution at the X-ray peak, approximately 1.5~keV FWHM.  This discrepancy potentially hints at a position-dependent energy collection efficiency in the detector, and we are currently studying the nature of this broadened peak. Finally, the source location in run A provided an $^{55}$Fe peak in the TES chip hit branch.  We use this to identify that the TES chip itself has a RMS resolution of $(12.3~\pm~0.2~\text{stat.}~\pm~0.3~\text{syst.})$~eV, with statistical uncertainty from the fit and the systematic uncertainty from the calibration.  Cuts for this resolution estimate were made in an identical way to the ``absorber hit" events, except we did not include the exponential amplitude cut.

\section{Conclusion}\label{conclusion}

We have presented the performance of the first iteration of a Q-Array-style \Ricochet detector.  These first results highlight many areas of further research, including studies of detector modeling and any potential position dependent-efficiencies.  Simultaneously, the baseline energy resolution achieved in this first iteration is a promising step towards fulfilling that which is described in the \Ricochet design requirements~\cite{augierFastNeutronBackground2023}.  

\section*{Acknowledgements}

This project has received funding from the NSF under Grants PHY-2209585 and PHY-2013203.  A portion of the work carried out at MIT was supported by DOE QuantISED award DE-SC0020181, the NSF under Grant PHY-2110569, and the Heising-Simons Foundation.  Work performed at Argonne National Laboratory and the Center for Nanoscale Materials, a U.S. Department of Energy Office of Science User Facility, was supported by the U.S. DOE, Offices of Basic Energy Sciences and High Energy Physics, under Contract No. DE-AC02-06CH11357.  This work also made use of the NUFAB facility of Northwestern University’s NUANCE Center, which has received support from the SHyNE Resource (NSF ECCS-2025633), the IIN, and Northwestern’s MRSEC program (NSF DMR-1720139).  Additionally, this project received funding from the European Research Council (ERC) under the European Union’s Horizon 2020 research and innovation program under Grant Agreement ERC-StG-CENNS 803079, the French National Research Agency (ANR) within the project ANR-20-CE31-0006, and the LabEx Lyon Institute of Origins (ANR-10-LABX-0066) of the Université de Lyon, within the Plan France2030.  This work is also partly supported by the Ministry of Science and Higher Education of the Russian Federation.


\bibliography{bibtex_jan12_short_authors}

\begin{thebibliography}{10}
\expandafter\ifx\csname url\endcsname\relax
  \def\url#1{\texttt{#1}}\fi
\expandafter\ifx\csname urlprefix\endcsname\relax\def\urlprefix{URL }\fi
\expandafter\ifx\csname href\endcsname\relax
  \def\href#1#2{#2} \def\path#1{#1}\fi

\bibitem{augierFastNeutronBackground2023}
{Ricochet Collaboration}, Fast neutron background characterization of the
  future {{Ricochet}} experiment at the {{ILL}} research nuclear reactor, The
  European Physical Journal C 83~(1) (2023) 20.
\newblock \href {https://doi.org/10.1140/epjc/s10052-022-11150-x}
  {\path{doi:10.1140/epjc/s10052-022-11150-x}}.

\bibitem{coherentcollaborationObservationCoherentElastic2017}
{Coherent Collaboration}, Observation of coherent elastic neutrino-nucleus
  scattering, Science 357~(6356) (2017) 1123--1126.
\newblock \href {https://doi.org/10.1126/science.aao0990}
  {\path{doi:10.1126/science.aao0990}}.

\bibitem{armengaudPerformanceEDELWEISSIIIExperiment2017}
E.~Armengaud, Q.~Arnaud, C.~Augier, et~al., Performance of the
  {{EDELWEISS-III}} experiment for direct dark matter searches, Journal of
  Instrumentation 12~(08) (2017) P08010.
\newblock \href {https://doi.org/10.1088/1748-0221/12/08/P08010}
  {\path{doi:10.1088/1748-0221/12/08/P08010}}.

\bibitem{edelweisscollaborationOptimizingEDELWEISSDetectors2018}
{EDELWEISS Collaboration}, Q.~Arnaud, E.~Armengaud, C.~Augier, et~al.,
  Optimizing {{EDELWEISS}} detectors for low-mass {{WIMP}} searches, Physical
  Review D 97~(2) (2018) 022003.
\newblock \href {https://doi.org/10.1103/PhysRevD.97.022003}
  {\path{doi:10.1103/PhysRevD.97.022003}}.

\bibitem{edelweisscollaborationFirstGermaniumBasedConstraints2020}
{EDELWEISS Collaboration}, Q.~Arnaud, E.~Armengaud, C.~Augier, et~al., First
  {{Germanium-Based Constraints}} on {{Sub-MeV Dark Matter}} with the
  {{EDELWEISS Experiment}}, Physical Review Letters 125~(14) (2020) 141301.
\newblock \href {https://doi.org/10.1103/PhysRevLett.125.141301}
  {\path{doi:10.1103/PhysRevLett.125.141301}}.

\bibitem{billardSearchingDarkMatter2021}
J.~Billard, Searching for {{Dark Matter}} and {{New Physics}} in the
  {{Neutrino}} sector with {{Cryogenic}} detectors, {Habilitation à Diriger
  des Recherches }{\url{https://theses.hal.science.tel-03259707}}, Universit\'e
  Claude Bernard Lyon 1 (Jan. 2021).

\bibitem{salagnacOptimizationPerformanceCryoCube2021}
T.~Salagnac, J.~Billard, J.~Colas, et~al., {Ricochet Collaboration},
  Optimization and performance of the {{CryoCube}} detector for the future
  {{RICOCHET}} low-energy neutrino experiment (Nov. 2021).
\newblock \href {http://arxiv.org/abs/2111.12438} {\path{arXiv:2111.12438}},
  \href {https://doi.org/10.48550/arXiv.2111.12438}
  {\path{doi:10.48550/arXiv.2111.12438}}.

\bibitem{colasDevelopmentDataProcessing2022}
J.~Colas, J.~Billard, S.~Ferriol, et~al., {for~the RICOCHET~collaboration},
  Development of {{Data Processing}} and {{Analysis Pipeline}} for the
  {{Ricochet Experiment}}, Journal of Low Temperature Physics (Nov. 2022).
\newblock \href {https://doi.org/10.1007/s10909-022-02907-5}
  {\path{doi:10.1007/s10909-022-02907-5}}.

\bibitem{irwinTransitionEdgeSensors2005}
K.~Irwin, G.~Hilton, Transition-{{Edge Sensors}}, in: C.~Enss (Ed.), Cryogenic
  {{Particle Detection}}, Topics in {{Applied Physics}}, {Springer}, {Berlin,
  Heidelberg}, 2005, pp. 63--150.
\newblock \href {https://doi.org/10.1007/10933596_3}
  {\path{doi:10.1007/10933596_3}}.

\bibitem{ricochetcollaborationRicochetProgressStatus2021}
{Ricochet Collaboration}, Ricochet {{Progress}} and {{Status}} (Nov. 2021).
\newblock \href {http://arxiv.org/abs/2111.06745} {\path{arXiv:2111.06745}},
  \href {https://doi.org/10.48550/arXiv.2111.06745}
  {\path{doi:10.48550/arXiv.2111.06745}}.

\bibitem{bastidonOptimizingThermalDetectors2018}
N.~Bastidon, J.~Billard, E.~{Figueroa-Feliciano}, et~al., Optimizing {{Thermal
  Detectors}} for {{Low-Threshold Applications}} in {{Neutrino}} and {{Dark
  Matter Experiments}}, Journal of Low Temperature Physics 193~(5) (2018)
  1206--1213.
\newblock \href {https://doi.org/10.1007/s10909-018-2073-2}
  {\path{doi:10.1007/s10909-018-2073-2}}.

\bibitem{chenTransitionEdgeSensor2022}
R.~Chen, H.~D. Pinckney, E.~{Figueroa-Feliciano}, et~al., Transition {{Edge
  Sensor Chip Design}} of a {{Modular CE}}ν{{NS Detector}} for the {{Ricochet
  Experiment}}, Journal of Low Temperature Physics (Dec. 2022).
\newblock \href {https://doi.org/10.1007/s10909-022-02927-1}
  {\path{doi:10.1007/s10909-022-02927-1}}.

\bibitem{angloherFirstMeasurementsRemoTES2023}
G.~Angloher, M.~R. Bharadwaj, I.~Dafinei, et~al., First measurements of
  {{remoTES}} cryogenic calorimeters: {{Easy-to-fabricate}} particle detectors
  for a wide choice of target materials, Nuclear Instruments and Methods in
  Physics Research Section A: Accelerators, Spectrometers, Detectors and
  Associated Equipment 1045 (2023) 167532.
\newblock \href {https://doi.org/10.1016/j.nima.2022.167532}
  {\path{doi:10.1016/j.nima.2022.167532}}.

\bibitem{gaitskellNonequilibriumSuperconductivityNiobium1993}
R.~Gaitskell, Non-equilibrium superconductivity in niobium and its application
  to particle detection, {PhD Thesis Chapter 7}
  {\url{https://ora.ox.ac.uk/objects/uuid:68020959-e878-4cd0-89c0-2d33c6b54013}},
  University of Oxford (1993).

\bibitem{2016JLTP..184...66L}
D.~{Li}, J.~E. {Austermann}, J.~A. {Beall}, et~al., {AlMn Transition Edge
  Sensors for Advanced ACTPol}, Journal of Low Temperature Physics 184~(1-2)
  (2016) 66--73.
\newblock \href {https://doi.org/10.1007/s10909-016-1526-8}
  {\path{doi:10.1007/s10909-016-1526-8}}.

\bibitem{pinckneyThermalConductanceSapphire2022}
H.~D. Pinckney, G.~Yacteen, A.~Serafin, et~al., The {{Thermal Conductance}} of
  {{Sapphire Ball Based Detector Clamps}}, Journal of Low Temperature Physics
  209~(5) (2022) 1204--1211.
\newblock \href {https://doi.org/10.1007/s10909-022-02777-x}
  {\path{doi:10.1007/s10909-022-02777-x}}.

\bibitem{MatterMethodsLow}
Matter and {{Methods}} at {{Low Temperatures}} | {{Frank Pobell}} |
  {{Springer}}, \url{https://www.springer.com/gp/book/9783540463566} (2007).

\bibitem{Pytesdaq2022}
Pytesdaq, SPICE/HeRALD GitHub \url{https://github.com/spice-herald/pytesdaq}
  (Apr. 2022).

\bibitem{QETpy2023}
The {IterCut} class from {QETpy} version 1.4.0, SPICE/HeRALD GitHub
  \url{https://github.com/spice-herald/QETpy} (Apr. 2023).

\end{thebibliography}

\end{document}